# Manganite-based three level memristive devices with self-healing capability

W. Román Acevedo[1,2], D. Rubi[1,2,3,*], J. Lecourt[4], U. Lüders[4], F. Gomez-Marlasca[1], P. Granell[5], F. Golmar[2,3,5] and P. Levy[1,2]

1. Gerencia de Investigación y Aplicaciones, CNEA, Av. Gral Paz 1499 (1650), San Martín, Buenos Aires, Argentina

2. Consejo Nacional de Investigaciones Científicas y Técnicas (CONICET), Argentina.

3. Escuela de Ciencia y Tecnología, UNSAM, Campus Miguelete (1650), San Martín, Buenos Aires, Argentina

4. CRISMAT, CNRS UMR 6508, ENSICAEN, 6 Boulevard Maréchal Juin, 14050 Caen Cedex 4, France

5. INTI - CMNB, Av. Gral Paz 5445 (B1650KNA), San Martín, Buenos Aires, Argentina

We report on non-volatile memory devices based on multifunctional manganites. The electric field induced resistive switching of $Ti/La_{1/3}Ca_{2/3}MnO_3/n\text{-}Si$ devices is explored using different measurement protocols. We show that using current as the electrical stimulus (instead of standard voltage-controlled protocols) improves the electrical performance of our devices and unveils an intermediate resistance state. We observe three discrete resistance levels (low, intermediate and high), which can be set either by the application of current-voltage ramps or by means of single pulses. These states exhibit retention and endurance capabilities exceeding $10^4$ s and 70 cycles, respectively. We rationalize our experimental observations by proposing a mixed scenario were a metallic filament and a $SiO_x$ layer coexist, accounting for the observed resistive switching. Overall electrode area dependence and temperature dependent resistance measurements support our scenario. After device failure takes place, the system can be turned functional again by heating up to low temperature (120ºC), a feature that could be exploited for the design of memristive devices with self-healing functionality. These results give insight into the existence of multiple resistive switching mechanisms in manganite-based memristive systems and provide strategies for controlling them.

\* Corresponding author (D.R.)

Tel: +54 11 67727059, Fax: +54 11 67727121, email: rubi@tandar.cnea.gov.ar

1. Introduction

The resistive switching (RS) mechanism displayed by certain metal – insulator – metal structures supports a new non-volatile memory technology coined as ReRAM [1-4]. For niche applications, ReRAM is a rival of other emerging replacement technologies of presently prevailing FLASH based RAM memory devices. Strong efforts have been made to fabricate structures that could exhibit RS. Typically, a metal–insulator–metal capacitor-like structure is assessed. Although many systems and materials were tested, the description of a unified physical mechanism is still a matter of debate [1-4]. Electric field induced filament formation and oxygen vacancy drift are key ingredients in most descriptions involving transition metal oxides as the insulating layer [3]. Here we focus on a manganese oxide multifunctional compound, $La_{1/3}Ca_{2/3}MnO_3$ (LCMO), a unique manganite well known for its colossal magnetoresistance and intrinsic phase separation effects [5]. Manganite based RS non-volatile memory devices were shown to exhibit unique properties, either in polycrystalline [6] or thin film [7,8] format. Electronic transport in manganites is described by electron hopping between neighbor Mn sites. As this hopping is mediated by oxygen ions existing between Mn atoms, the mechanism is known as "double exchange". Oxygen vacancies ($OV^+$) disrupt this hopping process, and thus increase the manganite electric resistance. On the other hand, the electrode material plays a key role, determining overall resistance levels by the introduction of a naturally formed oxide layer [9] or through the migration of metallic ions (i.e. metallic filament formation) [8,10]. Besides, manganite based non-volatile memory devices exhibit multilevel capability [11] sustained on the controlled electric field drift of vacancies [12], with the ability to tune (almost) continuously the actual resistance level. Here we perform current-controlled experiments that show that Ti/LCMO/n-Si devices exhibit, in addition to these multilevel (analogic) levels, three robust non-volatile (discrete) memory levels. In addition to the usual high and low resistance states, an intermediate resistance state can be stabilized. Unveiling of this three level memory device is attained through careful control of the dissipated electrical power. We elaborate on the nature of the observed RS mechanism, and propose a scenario where the experimental behavior is consistent with the existence of a Ti filament plus the oxidation/reduction of the ultrathin native $SiO_x$ layer present at the Si-manganite interface. Finally, we show that, after device failure takes place, our devices can be turned functional again by heating them at low temperatures. The possibility of taking advantage of this fact for the design of self-healing memristive devices is discussed.

2. Materials and Methods

LCMO manganite memory devices were grown on top of n-type silicon by pulsed laser deposition. No chemical removal of the native $SiO_x$ layer was performed. A 266nm Nd:YAG solid state laser, operating at a repetition frequency of 10Hz, was used. The deposition temperature and oxygen pressure were 850ºC and 0.13mbar, respectively. X-ray diffraction was performed by means of an Empyrean (Panalytical) diffractometer. Samples were single phase and polycrystalline, as shown in the X-ray diffraction pattern of Figure 1(a). The manganite peaks were indexed assuming a *Pnma* orthorombic bulk-like structure. The film thickness was estimated by focused ion beam cross-sectioning and scanning electron microscopy imaging in 100nm. The scanning electron microscopy plane view displayed in Figure 1(b) shows a dense microstructure along with the presence of a small amount of particulate on the surface, which is common in pulsed laser deposited films [13]. 100nm Ti top electrodes were deposited by sputtering and shaped by means of optical lithography. Top electrode areas ranged between $32\times10^3$ µm$^2$ and $196\times10^3$ µm$^2$. Electrical characterization was performed at room temperature with a Keithley 2612 source-meter hooked to a home-made probe station. The n-type silicon substrate was grounded and used as bottom electrode. The electrical stimulus was applied to the top electrode.

3. Results and discussion

Electroforming polarity and strength determines most of the subsequent electrical behavior of a metal – insulator – metal structure. In a previous work on similar Ti-LCMO-nSi structures, we reported that different transport mechanisms are triggered by either positive ($F^+$) or negative ($F^-$) electroforming [14]. In brief, $F^-$ determines the ulterior oxidation/reduction of the SiO$_x$ layer at the manganite –Si interface, while $F^+$ can be ascribed to either oxygen vacancies drift or to a metallic filament formation. Changes in the I-V curve circulation (clockwise vs. anticlockwise) with the electroforming polarity were also reported in SrTiO$_3$ and Ga$_2$O$_3$-based systems [15,16]; however, the physical mechanisms in these cases (related to the formation of oxygen vacancies paths, which locally lower the resistance of the oxide), are substantially different from our case. Results to be discussed below were obtained applying a positive stimulus to pristine devices, namely the $F^+$ procedure, though using the less common approach of current injection.

We started the electric testing protocol by recording pulsed voltage-current curves. Current pulses with different amplitudes (0→I$_{MAX}$→-I$_{MIN}$→0) and 1ms time-width were applied. Consecutive pulses were separated by ~1 sec in order to allow the heat produced at the electrode to drain. We recall that we used the applied current as the stimulus, and voltage as the dependent variable. The rationale behind is the control of the local power release during the expected abrupt transition from the initial high resistance (HR) state to the low resistance (LR) one, the SET operation. When using the "voltage control mode" dissipated power release obeys $P_v = V^2/R$. While driving the device in "voltage control mode" current flow is usually limited by the apparatus internal current compliance [17,18], as a way to avoid sample damage due to the power overshoot occurring during the sudden decrease of R (i.e. the desired RS effect obtained through a SET operation). However, the ever limited time response of standard equipment can not avoid an unwanted overshoot. Strategies using an external device for preventing this damage include the use of a biased transistor [19] or a simple resistance [20] in series with the memory cell. An alternative strategy is the use of the "current control mode" [21], in SET operations. Thus, during a sudden decrease of R, the dissipated power is self limited as power release is governed by $P_I=I^2R$.

Initially, all tested devices received *positive* electrical stimulus. We will focus now on the behavior of a device with area ~ $45 \times 10^3$ μm$^2$. The positive electroforming process determines a sudden decrease of the resistance from the virgin ($R_V$ ~ $2 \times 10^6$ Ω) to the HR resistance state value ($R_H$ ~ $80 \times 10^3$ Ω). Next, V - I curves were performed. Upon applying a positive stimulus, the HR resistance state (R ~ $80 \times 10^3$ Ω) can be explored for I < $2 \times 10^{-3}$ A. A sharp HR to LR (SET) transition occurs for I ~ $+2 \times 10^{-3}$A, as depicted in Fig. 2(a). When negative stimulus is explored, a transition from LR to HR (RESET) is observed to occur around $-10 \times 10^{-3}$A, reaching the original HR state. By cycling the stimulus between $+17 \times 10^{-3}$A and $-17 \times 10^{-3}$A, the SET – RESET operation can be repeated several times. Both HR and LR states are stable for at least $10^4$ s (i.e. negligible decay is observed when tested using a low non-disturbing stimulus). The LR state resistance $R_L$ ~ $2 \times 10^2$ Ω determines a resistance ratio $r_{HL}=R_H/R_L$~400.

Interestingly, we observed that if we extend the range of positive electrical stimulus, yet *another* transition is found. As shown in Fig. 2(b) a sharp (RESET-type) transition from the LR to an intermediate (IR) state is obtained around $+15 \times 10^{-3}$A.

This IR state exhibits R~$6 \times 10^3$ Ω, and was also found to be stable for at least $10^4$ s. The IR/LR ratio resulted $r_{IL}$=30. On the contrary of the other resistive transitions, the transition from LR to IR presents an instability, determined by the back and forth transitions between these states, as shown in the inset of Figure 2(b). The RESET transition from the IR state to the (initial) HR state is only obtained for *negative* polarity, as depicted in Fig. 2(b).

We should remark that the RS behavior observed in our current controlled experiments is definitely more stable and repetitive that in the case of voltage controlled mode. For instance, we have observed that in the latter case both HR and LR states progressively degrade and after aproximately 10 cycles the devices displays no further switching. In the case of current controlled experiments we avoid the uncontrolled current overshoot during the SET process (until the compliance current is established) and this enlarges the reliability of the device up to at least 70 cycles, as shown in the inset of Figure 2(a). Moreover, we argue that the access to the IR state is attained because of the controlled dissipated power during the SET procedure.

We have also checked that it is possible to switch directly between LR, IR and HR states by using single pulses of the appropriate amplitude, as a practical device would need. Figure 3(a) shows the resistance switching between LR and HR by applying opposite polarity current pulses with different amplitudes. In all cases, the time-width of the pulses was a few ms. The transition from HR to LR is achieved by applying +2mA pulses, while the opposite transition is obtained after the application of -60mA single pulses. Figure 3(b) (corresponding to a $196 \times 10^3 \mu m^2$ device) shows that the transition from LR to IR is obtained after the application of 4-pulse trains with increasing amplitudes up to 40mA. Figures 3(c) and (d) show the statistical resistance distributions of the LR, HR and IR states (obtained from a sequence of pulsed switchings as that shown in Figure 3(b)) and the retentivity of the three states up to ~$10^4$ s, respectively.

We will discuss now on the possible RS mechanisms that can account for the observed behavior attained after $F^+$. One of them is related to the electrical field induced migration of oxygen vacancies at the metal-manganite interface [12], which we have shown to dominate in voltage controlled experiments at low SET compliance currents on similar samples [8]. A positive stimulus applied to the Ti top electrode would produce a migration of $OV^+$ species from the metal – oxide interface towards the "bulk" LCMO material. The depletion of $OV^+$ from the interface region would produce a local *decrease* of the resistance, as oxygen ions reaching the interface region reinforce the electronic transport through the double exchange mechanism. The RESET transition observed at negative stimulus would be related with the total recovery of $OV^+$ towards the top electrode region. We notice that in this scenario a transition from LR to IR at positive polarities is not expected.

A second possibility is related to the creation of a conductive metallic bridge between both electrodes, related to the electrical field induced migration of the top electrode metal [8,10]. The electric field induced Ti drift would produce a metallic filament which may percolate, rendering a *decrease* of the resistance of the device, associated to the observed SET at I ~ $+2 \times 10^{-3}$A. In this scenario, the RESET transition observed at negative stimulus would be related to the retraction of the Ti filament. Moreover, the transition from LR to IR with positive stimulus could be explained by the partial disruption of the filament due to Joule heating (unipolar behavior) when reaching a high enough current level (i.e. I~$15 \times 10^{-3}$ A). Besides, the unstable switching between LR and IR states would be the consequence of the competition for filamentary formation / disruption produced by Ti-ions drift / Joule heating at the weakened region, where the electric field reaches its highest level.

A third possible mechanism is related to the oxidation/reduction of a $TiO_x$ layer formed at the Ti/manganite interface, as reported by Herpers et al. in Ref. [9]. However, in this case it is expected a SET (RESET) process with negative (positive) polarity, on the contrary of our experimental observation (Figure 2), indicating that this mechanism does not become dominant in our samples.

Further information about the involved RS mechanisms can be obtained by analyzing the behavior of the different resistance states as a function of the top electrode area. In the case of oxygen vacancies drift mechanism, a 1/Area dependence is expected for these resistance states as the electrical conduction is distributed all along the device area. On the other hand, a filamentary mechanism should not display any dependence of the LR with the device area, as in this case the electrical conduction is highly inhomogeneous and confined at the nanofilament area. Figure 4 depicts the area dependence of $R_H$, $R_I$ and $R_L$, obtained from several devices. Interestingly, neither the LR nor the IR states have strong area dependence, supporting the idea of filamentary based states.

However, in order to confirm the presence of a purely Ti-based filamentary mechanism an additional feature must be observed: the LR state should increase linearly as the temperature is raised, reflecting a metallic behavior [10]. This is in contrast with the measurements displayed in Figure 5, where it is observed that both LR and HR present a negative temperature coefficient, indicating an insulating behavior. This clearly shows that a purely (metallic) filamentary mechanism does not account for the observed behaviour of our system, and an additional ingredient should be taken into account.

A mixed scenario which combines evidences for filamentary behaviour on one hand, and semiconductor type transport on the other hand is envisaged. As strongly supported by F experiments, a possible scenario is related to the presence of the native $SiO_x$ layer at the manganite/Si interface. The thickness of this layer in Si substrates is usually around 1nm, although it can become thicker during the heating of the substrates in $O_2$ atmosphere prior to deposition. The electrical field induced oxidation and reduction of such an ultrathin layer was previously suggested to produce resistive switching [14,22]. Within this framework, we propose the following scenario, which is sketched in Figure 6. The LRS (SET process) is stabilized after the drift-diffusion of Ti ions coming from the top electrode through the manganite layer, forming a metallic nano-filament witch does not entirely connect both electrodes (a similar scenario was proposed in Ref. [23]). We notice that in the LRS the device resistance is dominated by the resistance of the oxides layers (LCMO and $SiO_x$) existing in between the filament tip and the n-Si electrode, explaining the semiconducting behavior observed in Figure 5(a). Upon the application of further positive stimulus, the high electrical field acting on the $SiO_x$ layer attracts negatively charged oxygen ions towards the $SiO_x$/manganite interface, resulting in a thinner but more stochiometric (and therefore more resistive) $SiO_{x+\delta}$ layer (we recall that oxygen non-stochiometries lowers the resistance of silicon oxide). This accounts for the the transition from LR to IR. The application of negative stimulus (RESET) produces the opposite process ($SiO_{x+\delta}$ → $SiO_x$) and also retracts the Ti filament, leading to the HR state. We recall that the fact that the IRS→HRS transition takes place in one single step implies that the (negative) current thresholds that trigger both (RESET) processes are similar.

A crosscheck was performed on Ti-LCMO-Pt structures, in which the manganite-$SiO_x$ layer does not exist. In such samples we found no evidence of the IR state. In contrast, a completely different transport scenario is observed as evidenced by the I-V curves, to be reported separately. Back to the Ti-LCMO-nSi structures, additional evidence for the presence of the $SiO_x$ layer is related to the observation that after device failure takes place (no further switching is found, the device resistance dropping to 30Ω), the device exhibits a resistance with a positive temperature coefficient (metallic behavior, see Figure 5(a)). This fact is likely related to the Ti filament penetrating into the $SiO_x$ until reaching the n-Si. Interestingly, we have found that if the device is heated up to 120ºC, the resistive switching behavior is recovered, as shown in Figure 5(b), indicating the local oxidation of (at least part of) the metallic filament with temperature. We also notice that if a Ti/LCMO device grown on platinized silicon is heated after electrical failure, no recovery of the RS is obtained, indicating that the Ti oxidation takes place locally for Ti atoms localized at the $SiO_x$ matrix. As the temperature level needed to reconstruct the device after failure (~100ºC) is relatively low, self heating could be produced by a proper design of the device (i.e. the geometrical dimension of the electrode) thus providing a "self healing" capability that could be activated through local power dissipation, consistently with the results recently reported by Tan et al. in $WO_{3-x}$ systems [24].

We would finally like to stress that the subtle electrical behavior we presented (which allows the stabilization of the intermediate IR state) is only seen in current controlled experiments, where the uncontrolled power release during the SET process is avoided. In voltage controlled experiments as those reported in [8], the $SiO_x$ is permanently broken during the SET procedure, and only two resistance states (LR, HR) are stabilized. An even subtler control of *both* SET and RESET processes can be achieved by using a combined strategy for the electrical stimulus if current control is used for the SET and voltage control is used for the RESET (a similar approach for the latter was used in Ref. [25]). In current control mode, the resistance increase during the RESET leads also to a power overshoot as $P=I^2R$; on the contrary, in voltage control mode $P=V^2/R$ and the power realeased during the RESET remains self-limited. Figure 5 shows a current-voltage sweep by applying this combined strategy, showing that a much more gradual RESET with multiple intemediate resistance levels is obtained.

4. Conclussions

In summary, we have shown that reliable three level memory states can be obtained on Ti/LCMO/n-Si resistive switching devices stimulated with controlled current. We attribute the observed behavior to a combination of Ti filaments formation together with the oxidation/reduction of the native silicon oxide layer at the manganite/Si interface. After device failure takes place, the system can be turned functional again by heating up to low temperatures (120ºC), suggesting an avenue for the design of memristive systems with self-healing capabilities. These results give insight into the presence of multiple resistive switching mechanisms in manganite-based memristive devices on silicon and provide strategies for controlling them.

Acknowledgements

We acknowledge financial support from CONICET (PIP 291), PICT "MeMO"(0788) and CIC-Buenos Aires. We thank Dr. D. Vega, from the Laboratory of X-ray Diffraction (GIA, GAIyANN, CAC, CNEA), for the XRD measurements.

# References

[1] A. Sawa, Mater. Today **11**, 28 (2008).

[2] R. Waser, R. Dittmann, G. Staikov and K. Szot, Adv. Mater. **21**, 2632 (2009)

[3] M. Rozenberg, Scholarpedia **6(4)**, 11414 (2011).

[4] D. B. Strukov and H. Kohlstedt, MRS Bulletin **37**, 108 (2012)

[5] Nanoscale Phase Separation and Colossal Magnetoresistance, E. Dagotto, Springer (2003).

[6] M. Quintero, A. G. Leyva, and P. Levy, Appl. Phys. Lett. **86**, 242102 (2005); M. Quintero, P. Levy, A. G. Leyva, and M. J. Rozenberg, Phys. Rev. Lett. **98**, 116601 (2007).

[7] Y.B. Nian, J. Strozier, N. J. Wu, X. Chen, and A. Ignatiev, Phys. Rev. Lett. **98**, 146403 (2007).

[8] D. Rubi, F. Tesler, I. Alposta, A. Kalstein, N. Ghenzi, F. Gomez-Marlasca, M. Rozenberg, and P. Levy, App. Phys. Lett. **103**, 163506 (2013).

[9] A. Herpers, C. Lenser, C. Park, F. Offi, F. Borgatti, G. Panaccione, S. Menzel, R. Waser and R. Dittmann, Adv. Mat. **26**, 2730 (2014)

[10] N. Ghenzi, M. J. Sánchez, D. Rubi, M. J. Rozenberg, C. Urdaniz, M. Weissman, and P. Levy, Appl. Phys. Lett **104**, 183505 (2014)

[11] N. Ghenzi, M. J. Sánchez, M. J. Rozenberg, P. Stoliar, F. G. Marlasca, D. Rubi and P. Levy, J. Appl. Phys. **111**, 084512 (2012).

[12] M. Rozenberg, M. J. Sánchez, R. Weht, C. Acha, F. G. Marlasca, and P. Levy, Phys. Rev. B **81**, 115101 (2010).

[13] Douglas B. Chrisey, Graham K. Hubler, Pulsed Laser Deposition on Thin Films, John Wiley & Sons, New York, 1994

[14] D. Rubi, A. Kalstein, W.S. Román, N. Ghenzi, C. Quinteros, E. Mangano, P. Granell, F. Golmar, F.G. Marlasca, S. Suarez, G. Bernardi, C. Albornoz, A.G. Leyva, P. Levy, Thin Solid Films **583**, 76 (2015).

[15] X.B. Yan, Y.D. Xia, H.N. Xu, X. Gao, H.T. Li, R. Li, J. Yin, and Z.G. Liu, Appl. Phys. Lett. **97**, 112101 (2010).

[16] X.B. Yan, H. Hao, Y.F. Chen, Y.C. Li, and W. Banerjee, Appl. Phys. Lett. **105**, 093502 (2014)]

[17] S. Yu, Y. Wu, R. Jeyasingh, D. Kuzu, and H.-S. Philip Wong, IEEE Trans. Elect. Dev. **58**, 2729 (2011).

[18] C. Rohde, B. Choi, D. Jeong, S. Choi, J. Zhao and C.S. Hwang, Appl. Phys. Lett. **86**, 262907 (2005).

[19] S.B. Lee, S.H. Chang, H.K. Yoo & Kang, J. Phys. D: Appl. Phys. **43**, 485103 (2010)

[20] D. Jana, S. Maikap, A. Prakash, Y.Y. Chen, H.C. Chiu and J.R. Yang, Nanoscale Res. Lett. **9**, 12 (2014)

[21] P. Mickel, A. John and M. Marinella, Appl. Phys. Lett. **105**, 053503 (2014).

[22] C. Li, H. Jiang and Q. Xia, Appl. Phys. Lett. **103**, 062104 (2013).

[23] X.L. Jiang, Y.G. Zhao, Y.S. Chen, D. Li, Y,X, Luo, D.Y. Zhao, Z. Sun, J.R. Sun, and H.W. Zhao, Appl. Phys. Lett. **102**, 253507 (2013)]

[24] Zheng-Hua Tan, Rui Yang, Kazuya Terabe, Xue-Bing Yin and Xin Guo, Phys. Chem. Chem. Phys. **18**, 1392 (2016)

[25] F. Yuang, Z. Zhang, L. Pan and J. Xu, J. Elect. Dev. Soc. **2**, 154 (2014).

**Figure Captions**

Figure 1: (a) X-ray diffraction pattern corresponding to one of our LCMO thin films. The inset shows a sketch of the device electrical connection; (b) Scanning electron micrograph corresponding to the same film.

Figure 2: (a) Room temperature V–I transport curves for a Ti/LCMO /n-Si device with area=$45\times10^3$ $\mu m^2$, exhibiting sweeps between $+10\times10^{-3}$A and $-10\times10^{-3}$A. The inset shows an endurance test consisting in 70 consecutive cycles; (b) V – I transport curves for the same device, with an extended I sweep range. The inset displays the electrical unstability associated to the transition between LR and IR states.

Figure 3: (a) Switching between LR and HR by applying single current pulses to a $32\times10^3$ $\mu m^2$ device; (b) Switching between LR, IR and HR by applying single pulses to a $196\times10^3$ $\mu m^2$ device. The different areas of both devices explain the difference in the HR state; (c) Statistical resistance distribution of the 3 resistance states; (d) Retentivity of the 3 resistance states up to $\sim10^4$ s.

Figure 4: (a) Remanent resistance for the LR, IR and HR states as a function of the electrode area.

Figure 5: (a) Evolution of LR and HR states as a function of the temperature. The behavior of a device after failure is also shown; (b) V-I curves corresponding to a device after failure and further heating up to 100ºC. The RS behavior is recovered after heating.

Figure 6: Sketch showing the proposed physical mechanisms behind each resistive state. A combination of Ti filament formation and oxidation and reduction of the native $SiO_x$ layer accounts for the experimental behavior.

Figure 7: V-I curve obtained by controlling current for positive bias and voltage for negative bias. A more subtle control of the RESET process is achieved with this strategy.

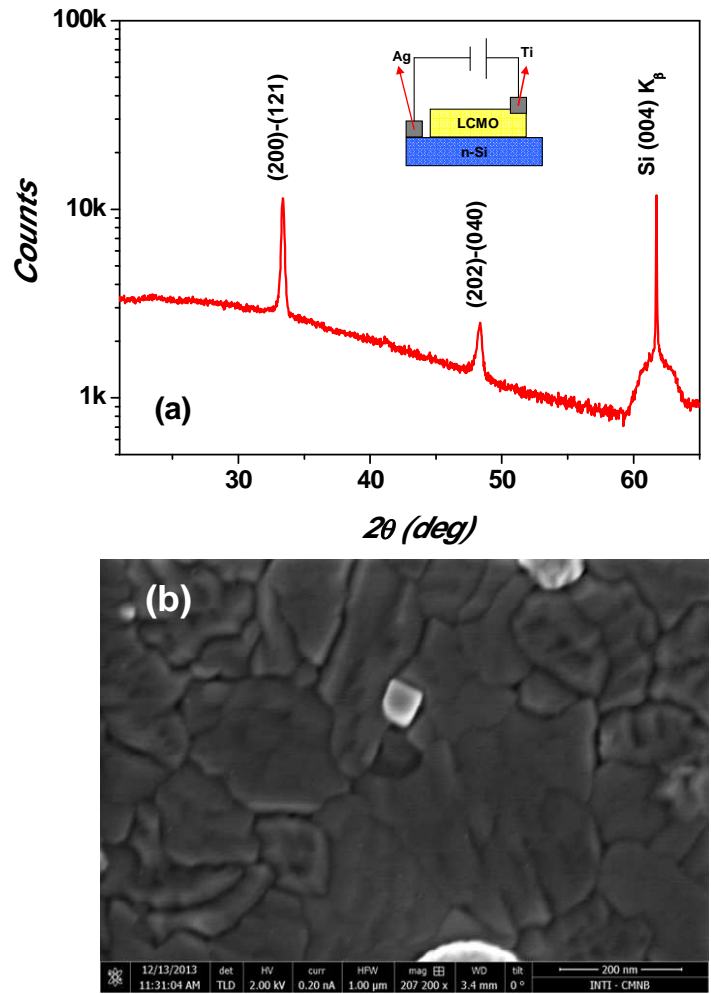

**FIGURE 1**

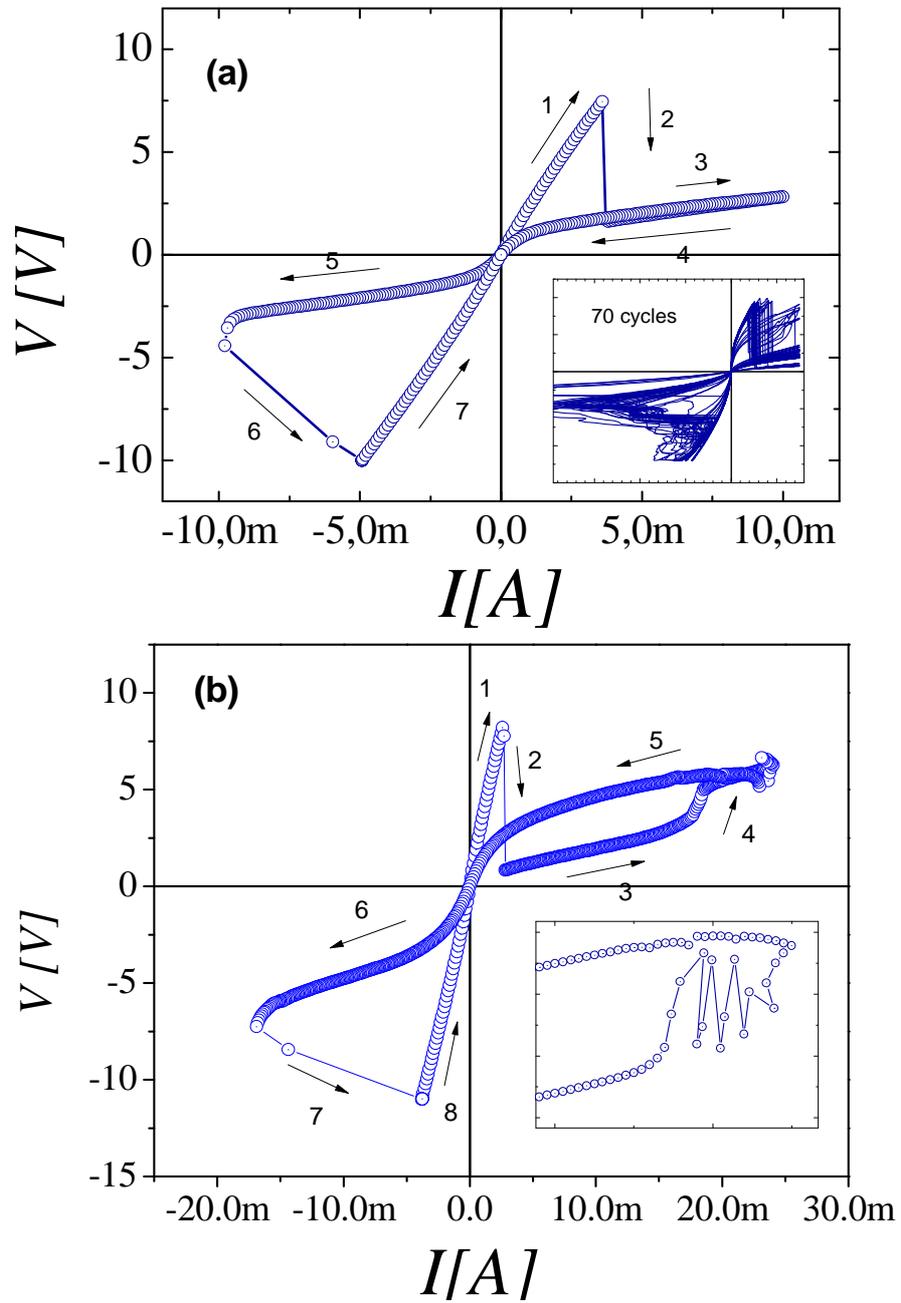

**FIGURE 2**

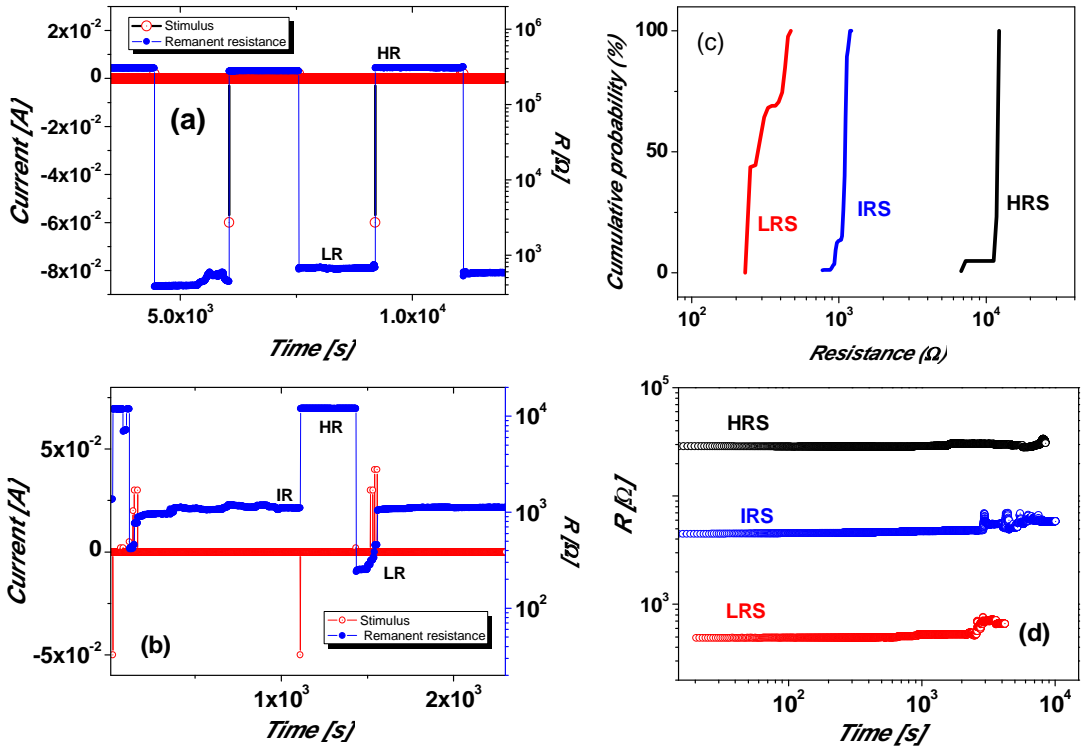

**FIGURE 3**

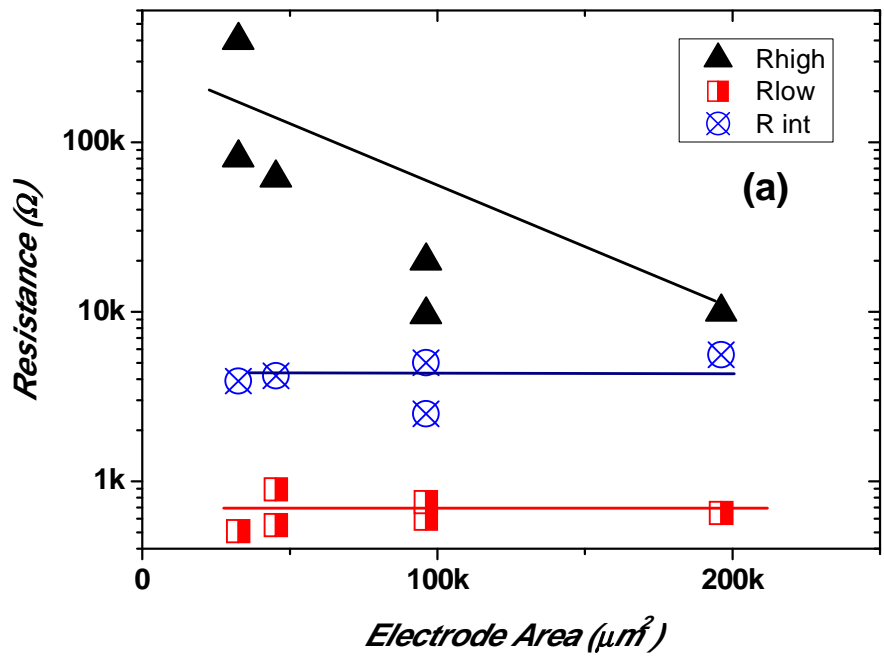

**FIGURE 4**

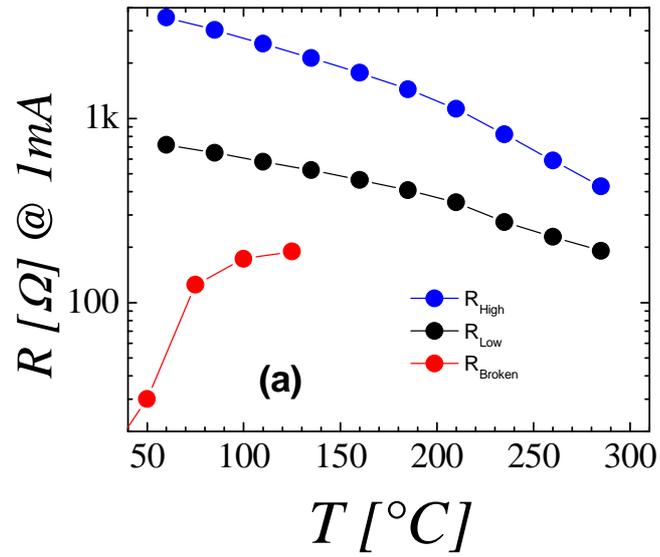

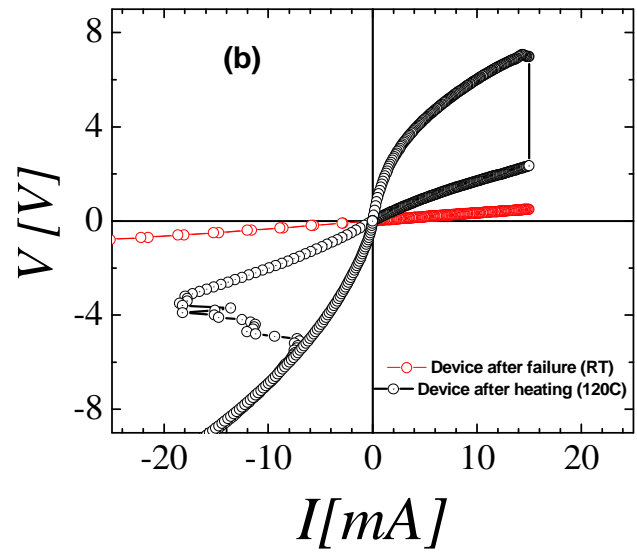

**FIGURE 5**

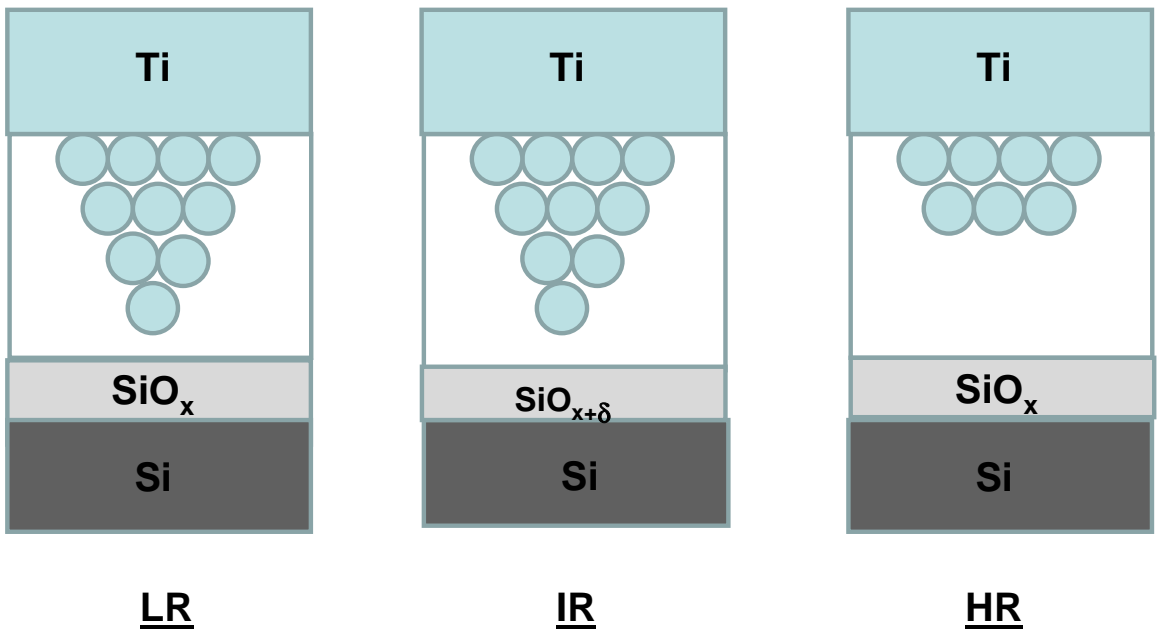

**FIGURE 6**

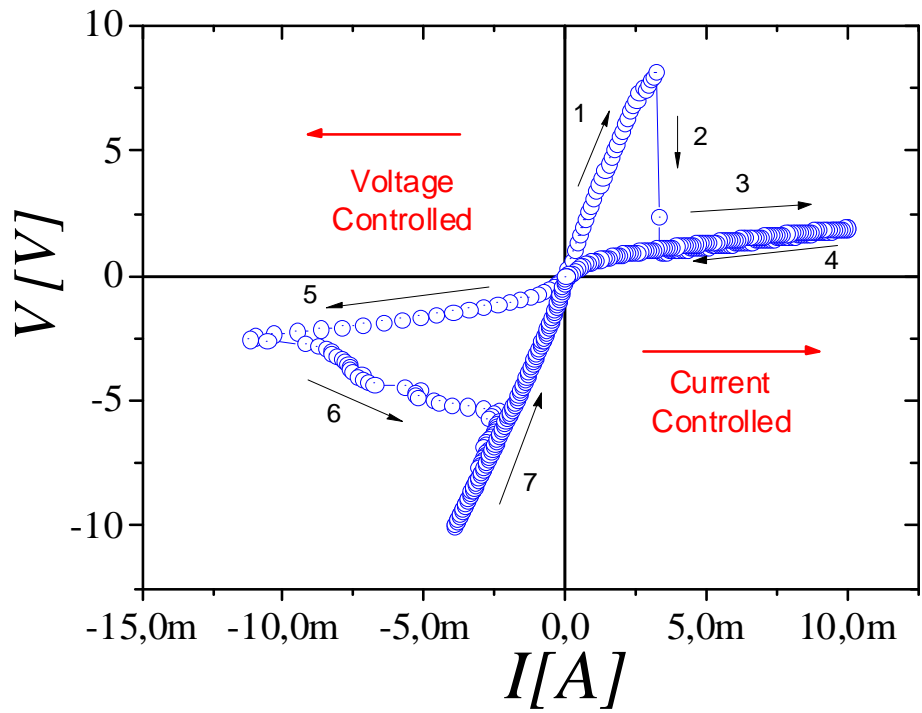

**FIGURE 7**